\documentclass[12pt]{iopart}

\usepackage{iopams}  
\usepackage{color}
\usepackage{graphicx}
\begin{document}

\title{``Plasma fibre" using bright-core helicon plasma}
\author{Lei Chang$^{1\ast}$, Zi-Chen Kan$^1$, Jing-Jing Ma$^1$, Saikat Chakraborty Thakur$^2$, and Juan Francisco Caneses$^3$}
\address{$^1$School of Electrical Engineering, Chongqing University, Chongqing, China}
\address{$^2$Department of Physics, Auburn University, Auburn, AL, United States}
\address{$^3$CompX, Del Mar, CA, United States}
\ead{leichang@cqu.edu.cn}

\begin{abstract}
This paper reports an innovative concept of ``plasma fibre" using bright-core helicon plasma, inspired by its spatial and spectral similarities to the well-known optical fibre. Theoretical analyses are presented for both ideal case of step-like density profile and the realistic case of Gaussian density profile in radius. The total reflection of electromagnetic waves near the sharp plasma density gradient and consequently the wave-guide feature could indeed happen if the incident angle is larger than a threshold value. Numerical computations using electromagnetic solver that based on Maxwell's equations and cold-plasma dielectric tensor yield consistent results. The experimental verification and prospective applications are also suggested. The ``plasma fibre" could be functional component that embedded into existing communication systems for special purpose based on its capability of dynamic reconfiguration.
\end{abstract}


\maketitle

\section{Background}\label{bgd}
\begin{figure}[ht]
\begin{center}
\includegraphics[width=0.85\textwidth,angle=0]{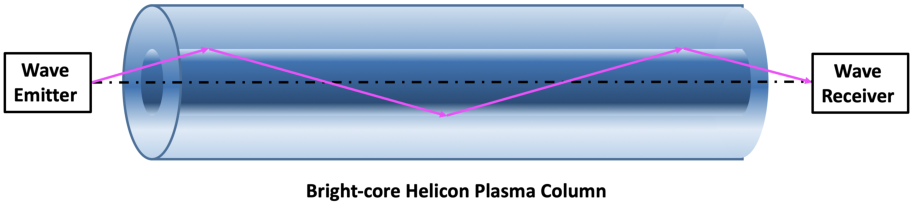}
\end{center}
\caption{Conceptual drawing of ``plasma fibre" using bright-core helicon plasma.}
\label{fg_drawing}
\end{figure}

The bright-core helicon plasma represents high-level helicon discharge that most plasma shrinks onto axis, forming a very thin and long cylinder, and emits dazzling bright colour (e.g. blue colour for argon) under certain experimental conditions (i.e. strong magnetic field and high RF power)\cite{Boswell:1984aa, Corr:2007aa, Blackwell:2012aa, Thakur:2015aa, Zhang:2021aa, Chang:2022aa}. This spatial structure of high density inside and low density outside, attributed to a transport barrier in radius that forming the edge of bright-core cylinder\cite{Thakur:2015aa, Thakur:2014aa}, resembles the structure of optical fibre if we consider the fact that the index of refraction is proportional to the plasma density\cite{Chen:2016aa, Stix:1992aa}. Indeed, the computation of wave field and power absorption shows preliminarily that the dispersive and dissipative features of bright-core plasma exhibit waveguide feature, similar to that of optical fibre\cite{Chang:2022ac}. In this work, we formally propose the innovative concept of ``plasma fibre" using bright-core helicon plasma. Figure~\ref{fg_drawing} displays the conceptual drawing. The analytical theory and numerical computations will be presented based on this.

\section{Theoretical analyses}\label{theory}
We consider two radial density profiles: step-like for ideal analysis, inspired by its similar structure to that of optical fibre, and Gaussian for realistic analysis, referring to actual bright-core helicon discharge experiments. Our motivation is trying to obtain the explicit condition of total reflection for wave propagation across the sharp density gradient. 

\subsection{Step-like density profile}
\begin{figure*}
\begin{center}$
\begin{array}{ll}
(a)&(b)\\
\includegraphics[width=0.68\textwidth,angle=0]{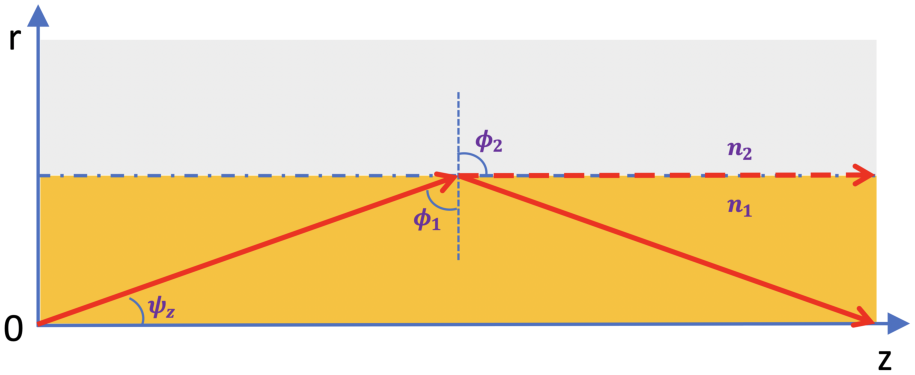}&\includegraphics[width=0.275\textwidth,angle=0]{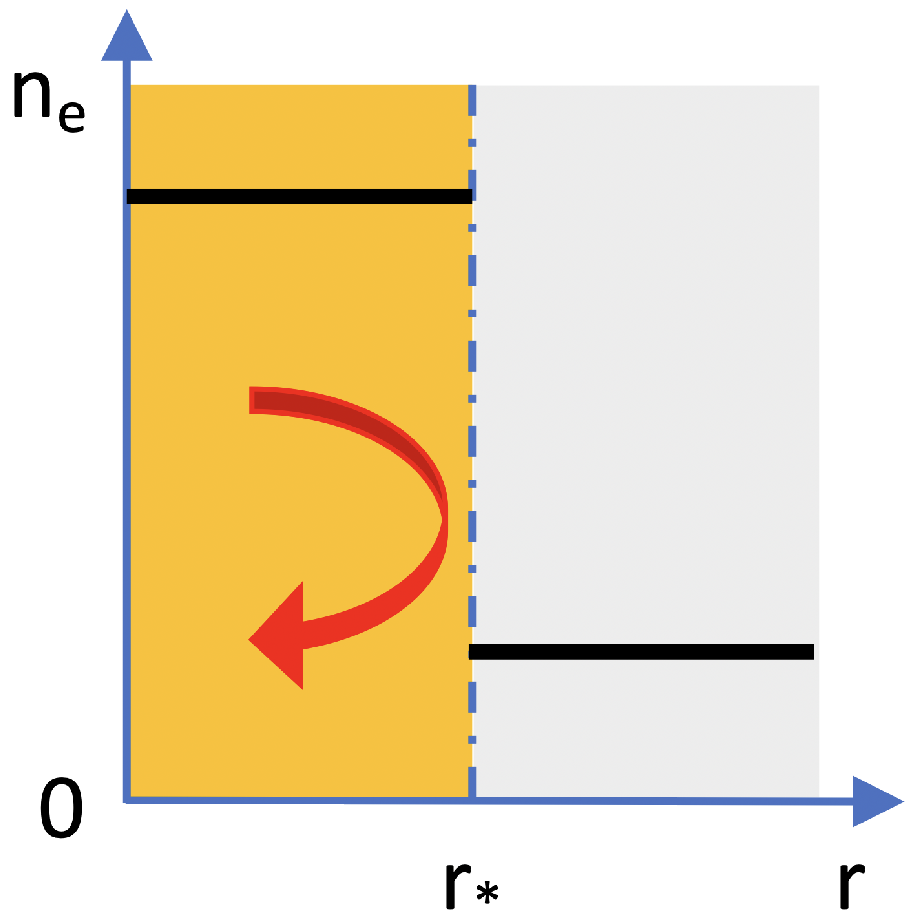}
\end{array}$
\end{center}
\caption{Illustration of the total reflection of wave propagation from step-like density gradient: (a) in $(r;~z)$ space, (b) in $(r;~n_e)$ space.}
\label{fg_steplike}
\end{figure*}

The total reflection of electromagnetic waves across a step-like density gradient in radius is very similar to that ocured in step-index fibre. An illustration is given in Fig.~\ref{fg_steplike}. The law of refraction, i.e. $n_1 \sin \phi_1=n_2\sin\phi_2$ with $n$ the index of refraction and $\phi$ the angle to normal direction, implies that the threshold angle for the total reflection ($\phi_2=\pi/2$) is $\phi_{1\ast}=\textrm{arcsin}~(n_2/n_1)$.
The index of refraction is defined as the ratio of the speed of light ($c$) and the phase velocity of waves ($v_{ph}$), i.e. $n=c/v_{ph}$ or equivalently $n=\lambda_0/\lambda$ with $\lambda_0$ the wavelength in vacuum and $\lambda$ inside the medium. As treated before\cite{Chang:2022ac}, for waves propagating in uniform magnetised plasma\cite{Ginzburg:1970aa, Gurnett:2005aa}, we can write $n$ as 
\small
\begin{equation}\label{eq1}
n^2=\frac{G\pm F}{2(\varepsilon \sin^2\psi_z+\eta \cos^2\psi_z)}
\end{equation}
\normalsize
with $\varepsilon$, g and $\eta$ the components of cold-plasma dielectric tensor, $\psi_z=\pi/2-\phi$, and:
\small
\begin{equation}\label{eq2}
G=(\varepsilon^2-g^2)\sin^2\psi_z+\varepsilon\eta(1+\cos^2\psi_z),
\end{equation}
\begin{equation}\label{eq3}
F^2=\left[(\varepsilon^2-g^2)-\varepsilon\eta\right]^2\sin^4\psi_z+4 g^2\eta^2\cos^2\psi_z.
\end{equation}
\normalsize
It is then straightforward to see that for either parallel waves:
\small
\begin{equation}\label{eq4}
n^2=1-\sum_\alpha\frac{\omega_{p\alpha}^2}{\omega(\omega\pm\omega_{c\alpha})}\approx \sum_\alpha\frac{\omega_{p\alpha}^2}{\omega(\omega\pm\omega_{c\alpha})}
\end{equation}
\normalsize
with $\omega$ the wave frequency, $\omega_{c\alpha}$ the cyclotron frequency, $\omega_{p\alpha}$ the plasma frequency, and the subscript $\alpha$ for ion or electron species, or oblique waves such as whistler mode:
\small
\begin{equation}\label{eq5}
n^2=\frac{\omega_{pe}^2}{\omega(\omega_{ce}\cos\psi-\omega)}, 
\end{equation}
\normalsize
the index of refraction is always proportional to the square root of plasma density, i.e. $n\propto \sqrt{n_{\alpha}}$ for high-density approximation ($\omega^2\ll \omega_{pe}^2$, $\omega_{c\alpha}^2\ll \omega_{pe}^2$), when other conditions are kept the same (wave frequency and magnetic field strength). Hence, the threshold value for total reflection could be written in form of
\small
\begin{equation}\label{eq6}
\phi_{1\ast}=\textrm{arcsin}~(\sqrt{n_{e\ast}}/\sqrt{n_{e0}})
\end{equation}
\normalsize
for step-like density profile in radius, with $n_{e\ast}$ and $n_{e0}$ the plasma densities outside and inside the jump radius ($r_\ast$), respectively.

\subsection{Continuous density profile} 
\begin{figure*}
\begin{center}$
\begin{array}{ll}
(a)&(b)\\
\includegraphics[width=0.68\textwidth,angle=0]{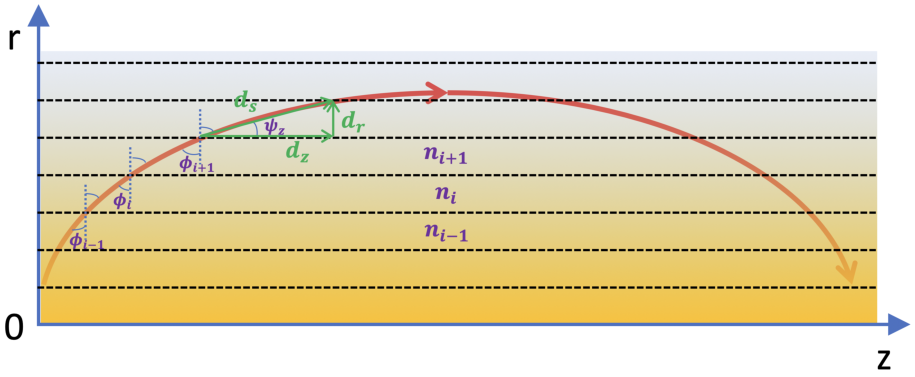}&\includegraphics[width=0.263\textwidth,angle=0]{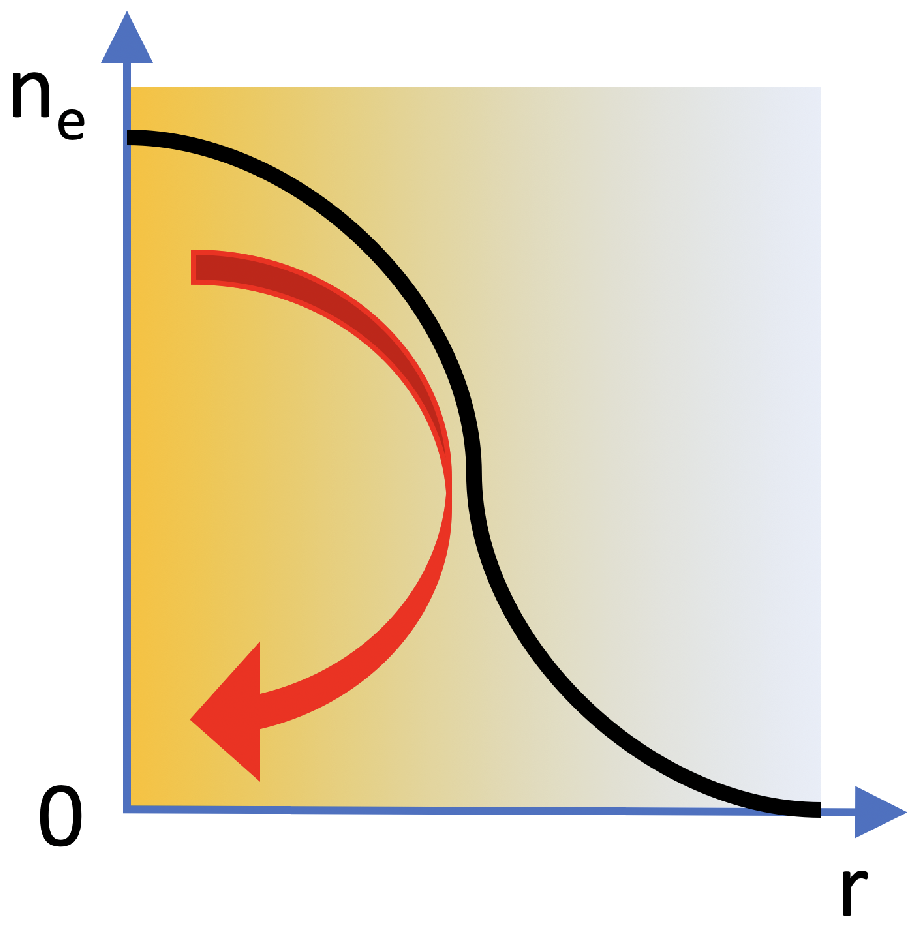}
\end{array}$
\end{center}
\caption{Illustration of the total reflection of wave propagation from Gaussian density gradient: (a) in $(r;~z)$ space, (b) in $(r;~n_e)$ space.}
\label{fg_continuous}
\end{figure*}

For real situation however the plasma density profile cannot be step-like but continuous in space, because of the diffusion and transport of particles. This results in continuously changing index of refraction and nonlinear path of wave propagation, i.e. the ray path is not a straight line but a curve. Figure~\ref{fg_continuous} shows an illustration. Different from the step-like case, there is no specific radius here for density jump and the total reflection does not occur at a specific layer but in a transitional area from core to edge and composes multiple small-angle refractions. 
To analyse it theoretically, we divide the radial domain into layers, inspired by the inner region of graded-index fibre. The number of the layers is so large that the ray path across neighbouring layers can be treated as a straight line and the index of refraction inside each layer can be taken as constant. Consequently, we can write $n_{i-1}\sin\phi_{i-1}=n_{i}\sin\phi_{i}=n_{i+1}\sin\phi_{i+1}$, or alternatively we could rewrite this relation in form of $n(r)\cos(\psi_z)=n_0\cos(\psi_{z0})$. The trigonometric function can be also expressed as $\cos(\psi_z)=dz/ds$ with $ds=\sqrt{dz^2+dr^2}$. Therefore, we obtain the following ray-path equation
\begin{equation}\label{eq7}
n(r)\frac{dz}{\sqrt{dz^2+dr^2}}=n_0\cos(\psi_{z0}),
\end{equation}
or equivalently
\begin{equation}\label{eq8}
dz=\frac{n_0\cos(\psi_{0z})}{\sqrt{n^2(r)-n_0^2\cos^2(\psi_{z0}})}dr.
\end{equation}
This determines the ray path of wave propagation in a plasma column with continuously changing index of refraction. We then set hyperbolic secant distribution for the index of refraction, i.e. $n(r)=n(0) \textrm{sech} (A r)$, in order to generate self-focusing characteristics and eliminate the phase distortion from emitter to receiver, as usually chosen for optical fibre. The integration of Eq.~(\ref{eq8}) results in
\begin{equation}\label{eq9}
z=\int \frac{n_0\cos(\psi_{0z})}{\sqrt{n^2(r)-n_0^2\cos^2(\psi_{z0}})}dr+C,
\end{equation}
which can be further written of the form
\begin{equation}\label{eq10}
z=\int \frac{\cosh(A r)}{\sqrt{\frac{n^2(0)-n_0^2\cos^2(\psi_{z0})}{n_0^2\cos^2(\psi_{z0})}-\sinh^2(A r)}}dr+C
\end{equation}
based on $\textrm{sech} (A r)=1/\cosh (A r)$, $\cosh^2 (A r)=1+\sinh^2 (A r)$. By employing the variable substitution of $x=\sinh(A r)$ (so that $dx=A\cosh(A r)dr$), we can further obtain
\begin{equation}\label{eq11}
z=\frac{1}{A}\arctan \left[\frac{n_0\cos\psi_{z0}\sinh (A r)}{\sqrt{n^2(0)-n_0^2\cos^2\psi_{z0}\cosh^2(A r)}}\right]+C,
\end{equation}
or equivalently
\begin{equation}\label{eq12}
\tan A(z-C)=\frac{n_0\cos\psi_{z0}\sinh (A r)}{\sqrt{n^2(0)-n_0^2\cos^2\psi_{z0}\cosh^2(A r)}}.
\end{equation}
The total reflection occurs where $\psi_{z0}=0$ so that $\cos\psi_{z0}=1$. This turns Eq.~(\ref{eq12}) into
\begin{equation}\label{eq13}
\tan A(z-C)=\frac{n_0\sinh (A r)}{\sqrt{n^2(0)-n_0^2\cosh^2(A r)}}.
\end{equation}

Equation~(\ref{eq6}) and Eq.~(\ref{eq13}) describe analytically the total reflection conditions of ``plasma fibre" using bright-core helicon plasma, and will compared qualitatively with following numerical computations. 

\section{Numerical computations}\label{cmp}
We shall use a well benchmarked electromagnetic solver (EMS)\cite{Chen:2006aa} to compute the wave field, dispersion relation, wave energy, and power absorption for wave propagations inside bright-core helicon plasma. The motivation is trying to characterise the waveguide features of bright-core helicon plasma, and compare them with those of the well-known optical fibre, together with the analytical calculations given above. The computational domain for EMS is constructed referring to the CSDX experiments\cite{Thakur:2015aa, Thakur:2014aa}. Two layouts of driving antenna are considered: outside plasma column (circulating quartz tube) and inside bright core, respectively. While the former is from experiment, the latter is excite waves near the axis and simulate the waveguide behaviour of bright-core helicon plasma. Other conditions are taken from the experiments directly and same to our previous study\cite{Chang:2022ac}. They include external magnetic field of $0.16$~T and electron temperature of $4$~eV, which are uniform throughout the whole domain. 

\begin{figure}[ht]
\begin{center}
\includegraphics[width=0.5\textwidth,angle=0]{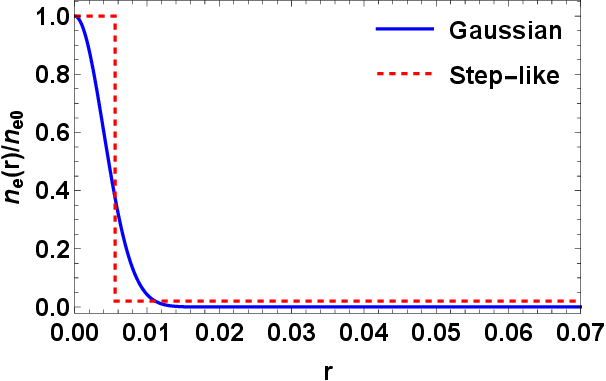}
\end{center}
\caption{Radial profiles of plasma density employed in the EMS simulations: ``Step-like" is constructed as ideal case, ``Gaussian" refers to actual CSDX device\cite{Thakur:2015aa, Thakur:2014aa, Burin:2005aa}.}
\label{fg_density}
\end{figure}

The utilised radial profiles of plasma density are shown in Fig.~\ref{fg_density}. They both peak on axis of the magnitude $4.72\times 10^{19}~\textrm{m}^{-3}$. Two configurations are considered particularly: step-like profile as an ideal case for comparing with optical fibre and the realistic case of Gaussian profile based on typical helicon discharges\cite{Boswell:1970aa, Boswell:1984aa, Blackwell:2012aa, Thakur:2015aa}. The location and magnitude of the density jump for the step-like profile are constructed as close as possible to the Gaussian profile, to ensure similar size of bright-core plasma column. We scanned a wide range of frequency, i.e. $0.1\sim 5$~GHz, to explore the wave propagation features of this bright-core helicon plasma. The following are typical results of computed wave field, while detailed evolutions in both radial and axial directions are manipulated into videos that available from the metadata repository of this paper. 

\subsection{Wave field structure and dispersion relation}
\begin{figure}[ht]
\begin{center}$
\begin{array}{ll}
(a)&(b)\\
\includegraphics[width=0.47\textwidth,angle=0]{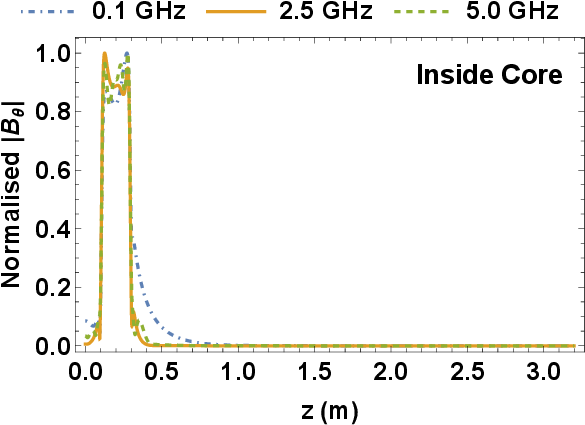}&\hspace{-0.15cm}\includegraphics[width=0.478\textwidth,angle=0]{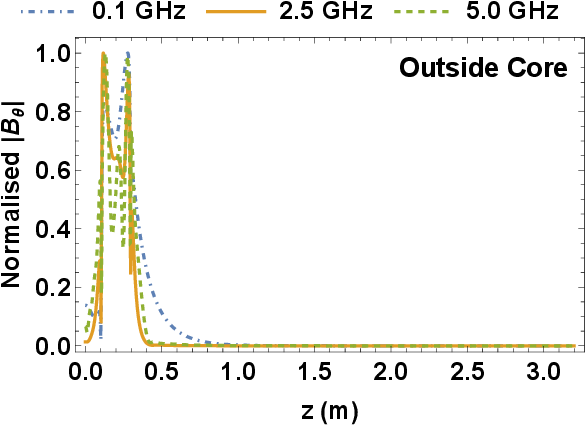}\\
(c)&(d)\\
\hspace{0.1cm}\includegraphics[width=0.475\textwidth,angle=0]{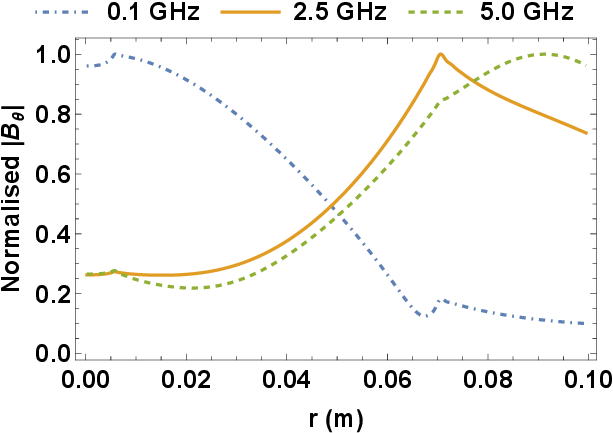}&\includegraphics[width=0.47\textwidth,angle=0]{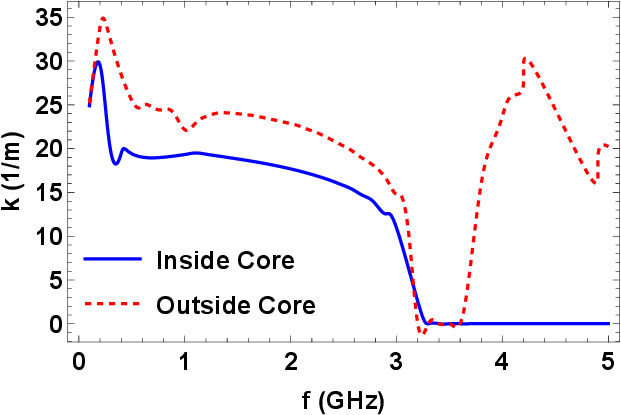}\\
\end{array}$
\end{center}
\caption{Typical results of computed wave field for the step-like density profile shown in Fig.~\ref{fg_density}: (a) axial variations inside core, (b) axial variation outside core, (c) radial variations across the core, (d) dispersion relations for frequency range of $0.1-5$~GHz.}
\label{fg_wf_stp}
\end{figure}

For step-like density profile, the typical results of computed wave field are given in Fig.~\ref{fg_wf_stp}.
From the axial variations measured inside the core ($r=0$~m) and outside the core ($r=0.016$~m), we can see that the wave field is mostly localised near the antenna and cannot propagate further towards the right end. Although the radial variations ($z=1$~m) change significantly when the frequency increases, the dispersion relation remains largely the same for $f\leq 3.1$~GHz, whereas the wave number outside the core is slightly bigger than that inside. However, for $f>3.1$~GHz, the wave propagation vanishes inside the core but boots up outside the core. This indicates that the plasma column with step-like density profile does act like a waveguide as constructed referring to optical fibre, namely the core can behave as band-pass filter with cut-off frequency. 

\begin{figure}[ht]
\begin{center}$
\begin{array}{ll}
(a)&(b)\\
\includegraphics[width=0.47\textwidth,angle=0]{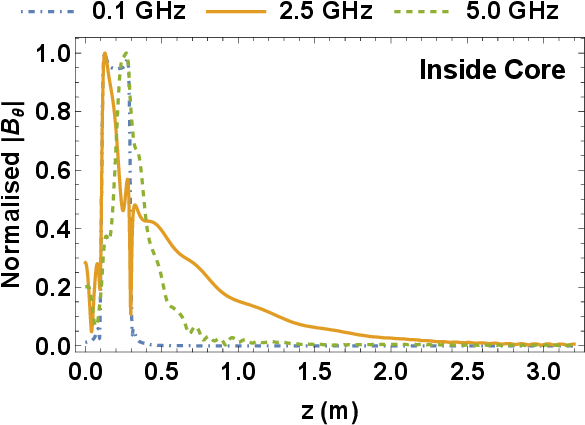}&\hspace{-0.15cm}\includegraphics[width=0.478\textwidth,angle=0]{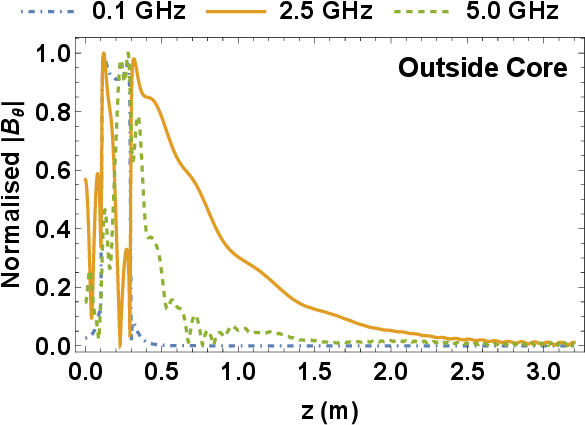}\\
(c)&(d)\\
\hspace{0.1cm}\includegraphics[width=0.475\textwidth,angle=0]{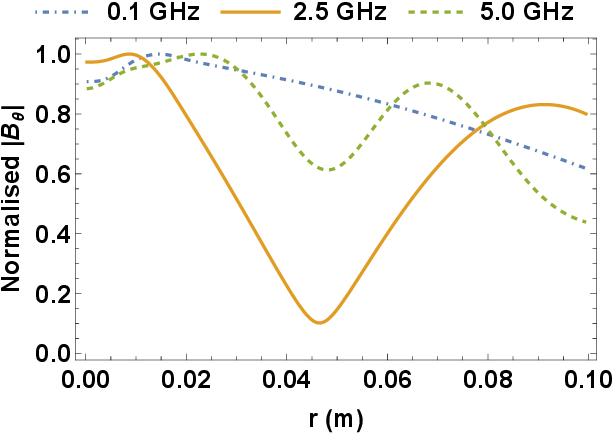}&\includegraphics[width=0.47\textwidth,angle=0]{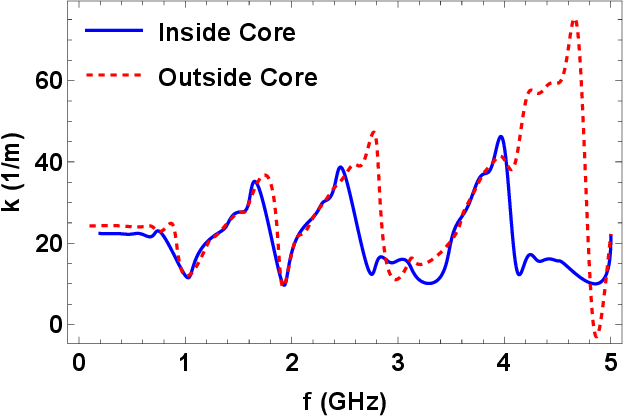}\\
\end{array}$
\end{center}
\caption{Typical results of computed wave field for Gaussian density profile shown in Fig.~\ref{fg_density}: (a) axial variations inside core, (b) axial variation outside core, (c) radial variations across the core, (d) dispersion relations for frequency range of $0.1-5$~GHz.}
\label{fg_wf_gsn}
\end{figure}

For Gaussian density profile, the typical results of computed wave field are presented in Fig.~\ref{fg_wf_gsn}.
Different from Fig.~\ref{fg_wf_stp}, the wave propagation is not localised near the antenna but can approach the right end for certain frequencies. We actually found two isolated frequency windows of $1.9-2.5$~GHz and $3.6-4.0$~GHz, respectively, for which the axial attenuations are relatively weak. These propagations are accompanied by multiple radial modes. This feature is consistent with the eigenmode behaviour of waveguide. The dispersion relations measured inside the core is similar to that outside, except for the high frequency range of $f>4$~GHz. Again, this looks like a waveguide, i.e. band-pass filter with cut-off frequency. 

\subsection{Frequency dependence and waveguide characteristics}
\begin{figure}[ht]
\begin{center}
\includegraphics[width=0.99\textwidth,angle=0]{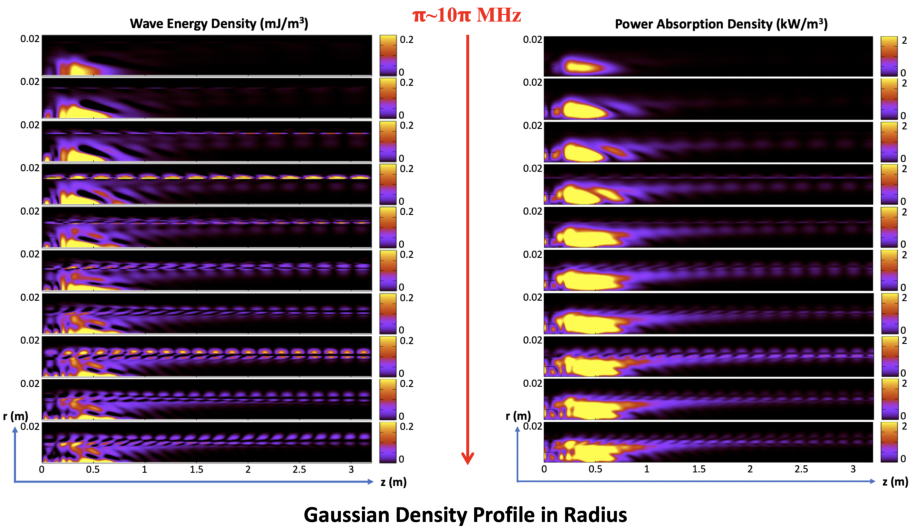}
\end{center}
\caption{2D wave energy density and power absorption density for frequency range of $\pi\sim 10\pi$~MHz with step size of $\pi$~MHz, showing waveguide feature inside bright-core plasma column.}
\label{fg_2d}
\end{figure}

To illustrate more clearly the waveguide features of bright-core helicon plasma, we plot the two-dimensional (2D) distributions of wave energy density and power absorption density for frequency range of $\pi\sim 10\pi$~MHz (with a step size of $\pi$~MHz), for example. Figure~\ref{fg_2d} shows the results. The plasma adopts Gaussian radial density profile, corresponding to the realistic configuration shown in Fig.~\ref{fg_density}. This figure provides a direct visualisation of how the electromagnetic waves evolve with frequency and demonstrates the emergence of waveguide behaviour inside the bright-core helicon plasma column.
At low frequencies ($f<3\pi$~MHz), the wave energy and power absorption are strongly localised near the antenna region, and only weak penetration along the axial direction is observed. This corresponds to the sub-cutoff regime, consistent with the analytical prediction that total reflection and guided propagation cannot be sustained when the refractive index contrast between core and edge is insufficient. As the frequency increases ($4\pi\leq f\leq 8\pi$~MHz), the energy density and power absorption contours extend significantly downstream, forming periodic axial structures and multiple radial lobes. These features are characteristic of guided eigenmodes in a cylindrical waveguide and are in excellent agreement with the multi-mode propagation predicted by the graded-index analysis in Eq.~(\ref{eq13}). The periodic modulation along $z$ further reflects the constructive interference of forward-propagating and weakly reflected waves within the bright-core region, signifying efficient energy confinement. In the higher-frequency regime ($f>8\pi$~MHz), the field distribution gradually shifts outward and the axial attenuation becomes more pronounced, indicating that the guiding condition weakens as the wavelength approaches the characteristic scale of the density gradient. This marks the onset of cut-off behaviour, similar to that observed in step-index optical fibres when the incident angle falls below the critical value defined in Eq.~(\ref{eq6}). The comparison between the left and right panels of Fig.~\ref{fg_2d} reveals that the power absorption follows closely the spatial distribution of the wave energy density, confirming that the bright-core helicon plasma supports effective energy transport along the axis within specific frequency windows. Overall, the results in Fig.~\ref{fg_2d} provide strong numerical evidence of waveguide-like confinement and frequency-selective propagation in a plasma column with Gaussian density profile. The findings are fully consistent with the theoretical conditions derived in Sec.~\ref{theory} and the simulation results presented in Fig.~\ref{fg_wf_stp} and Fig.~\ref{fg_wf_gsn}. They collectively validate the central concept of this work—the “plasma fibre”, which behaves as a dynamically reconfigurable electromagnetic waveguide sustained by plasma density gradients rather than solid boundaries.

\section{Summary}
This Letter reports a novel concept of ``plasma fibre” using bright-core helicon plasma, inspired by its strong spatial and spectral analogy to optical fibre. Through rigorous theoretical analysis and electromagnetic simulations based on Maxwell’s equations and the cold-plasma dielectric tensor, we demonstrate that helicon plasmas with steep radial density gradients can exhibit total internal reflection and waveguide characteristics similar to those of optical fibres. Both ideal step-like and realistic Gaussian density profiles are investigated, and the simulations reveal frequency-dependent propagation windows and cut-off behaviour indicative of guided-wave modes. These findings not only provide a new physical interpretation for wave propagation in bright-core helicon plasmas but also suggest potential applications of ``plasma fibres” as dynamically reconfigurable waveguides in communication or diagnostic systems. The work represents a significant conceptual advance that bridges plasma physics and photonics, offering broad interest to readers in plasma wave theory, plasma-material interaction, and advanced plasma applications.

To experimentally verify the ``plasma fibre” concept, we propose to generate a bright-core helicon discharge with a well-defined radial density gradient and to diagnose the associated electromagnetic wave propagation using movable B-dot and Langmuir probes. By scanning the driving frequency and magnetic field, the axial attenuation and radial localisation of the wave field can be mapped to identify frequency windows where wave energy is confined within the bright core and cut-off behaviour appears outside, as predicted by the simulations. Complementary optical imaging and emission spectroscopy can further reveal the correspondence between power absorption and wave energy density. Observation of sustained, core-localised propagation and frequency-selective confinement would provide direct experimental confirmation of the ``plasma fibre” as a self-organised plasma waveguide, thereby bridging the theoretical and numerical findings presented in this Letter.

\section{Supplementary Material}
Manipulated videos are available online, showing the detailed evolutions of wave field in both radial and axial directions as the driving frequency is varied. 

\ack
Discussions with Saikat Thakur, Juan Caneses, Jingchun Li, Huasheng Xie, Bo Li, and Chaofeng Sang are appreciated. This research is supported by National Natural Science Foundation of China (92271113, 12411540222, 12481540165), the Fundamental Research Funds for Central Universities (2022CDJQY-003), the Chongqing Entrepreneurship and Innovation Support Programme for Overseas Returnees (CX2022004), the Natural Science Foundation Project of Chongqing (CSTB2025NSCQ-GPX0725), and the ENN's Hydrogen-Boron Fusion Research Fund (Grant Number: 2025ENNHB01-011).

\section*{Data Availability Statement}
The data that support the findings of this study are available from the corresponding authors upon reasonable request.

\section*{ORCID IDs}
\raggedright
Lei Chang: https://orcid.org/0000-0003-2400-1836

Zi-Chen Kan: https://orcid.org/0009-0008-6506-7018

Jing-Jing Ma: https://orcid.org/0009-0004-1414-8273

\section*{References}

\bibliographystyle{unsrt}

\end{document}